\documentclass{optica-article}

\journal{opticajournal} 

\articletype{Research Article}

\usepackage{lineno}

\usepackage{subcaption}
\usepackage[utf8]{inputenc}
\begin{document}

\title{Real-Time Distributed Optical Fiber Vibration Recognition via Extreme Lightweight Model and Cross-Domain Distillation}

\author{Zhongyao Luo,\authormark{1} Hao Wu,\authormark{1,*} Zhao Ge, \authormark{1} and Ming Tang\authormark{1}}

\address{\authormark{1}Wuhan National Laboratory for Opto-electronics, Next Generation Internet Access National Engineering Laboratory, and Hubei Optics Valley Laboratory, School of Optical and Electronic Information, Huazhong University of Science and Technology, Wuhan, China}

\email{\authormark{*}wuhaoboom@hust.edu.cn} 


\begin{abstract} 
Distributed optical fiber vibration sensing (DVS) systems offer a promising solution for large-scale monitoring and intrusion event recognition. 
However, their practical deployment remains hindered by two major challenges: (1) degradation of recognition accuracy in dynamic conditions, and (2) the computational bottleneck of real-time processing for mass sensing data. 
This paper presents a new solution to these challenges, through a FPGA-accelerated extreme lightweight model along with a newly proposed knowledge distillation framework.
The proposed three-layer depthwise separable convolution network contains only 4141 parameters, which is the most compact architecture in this field to date, and achieves a maximum processing speed of 0.019 ms for each sample covering a 12.5 m fiber length over 0.256 s. This performance corresponds to real-time processing capabilities for sensing fibers extending up to 168.68 km.
To improve generalizability under changing environments, the proposed cross-domain distillation framework guided by physical priors is used here to embed frequency-domain insights into the time-domain model. This allows for time-frequency representation learning without increasing complexity and boosts recognition accuracy from 51.93\% to 95.72\% under unseen environmental conditions.
The proposed methodology provides key advancements including
a framework combining interpretable signal processing technique with deep learning and
a reference architecture for real-time processing and edge-computing in DVS systems, and more general distributed optical fiber sensing (DOFS) area. It mitigates the trade-off between sensing range and real-time capability, bridging the gap between theoretical capabilities and practical deployment requirements.
Furthermore, this work reveals a new direction for building more efficient, robust and explainable artificial intelligence systems for DOFS technologies.
\end{abstract}

\section{Introduction}
Optical fiber sensing (OFS) plays an increasingly important role in various fields, offering unique advantages, including immunity to electromagnetic interference, high sensitivity, robustness and flexibility\cite{pendaoOpticalFiberSensors2022}.
Distributed optical fiber sensors (DOFS) extends these benefits by transforming standard optical fibers into continuous sensing media, effectively creating kilometer-scale sensor arrays, through analysis of back-scattering patterns\cite{zhouHybridDistributedOptical2023}.
Among DOFS variants, distributed vibration sensing (DVS) based on phase-sensitive optical time-domain reflectometry (Phi-OTDR) has gained particular attention due to its exceptional spatial resolution and cost effectiveness in large-area monitoring applications.
This allows it to be applied in various fields including geotechnical engineering\cite{ouelletPreviouslyHiddenLandslide2024}, pipeline monitoring\cite{gietzMachineLearningAutomated2024, jiaPipelineLeaksEarly2024, lyuTwoStageIntrusionEvents2024}, railway monitoring\cite{xieRailwayTrackOnline2024, heDefectRecognitionBallastless2023}, structure health monitoring\cite{fengIdentificationCorrosionInducedOptical2024}, and traffic monitoring\cite{dingDistributedAcousticSensingBased2024}.

Modern DVS systems leverage Rayleigh back-scattering lights induced by narrow-linewidth laser pulses propagating through optical fibers. 
External vibrations alter the fiber's optical parameters, generating detectable phase and intensity variation in the back-scattered signal.
Analysis of these change patterns enables simultaneous vibration location and characterization over multi-kilometer ranges. 
However, practical implementation faces fundamental challenges, including complex environmental noises due to the variable environment the fiber might encounter, signal pattern variability due to medium heterogeneity and coupling effects, and large data volume from sensing data captured.

These technical constraints are compounded by strict operational requirements in time-sensitive applications such as pipeline monitoring and earthquake detection \cite{Hernndez2022DeepLearningBasedED}. 
Field deployments cannot be sure to provide abundant computation resources and power supply.
This necessitates DVS algorithms that simultaneously satisfy critical criteria including real-time processing of high-dimensional data streams, robustness to non-stationary signal characteristics, and deployability in resource-constrained environments.

Conventional signal processing approaches employ manually engineered feature extractors requiring extensive domain expertise for application-specific calibration. 
While deep learning approaches automate feature extraction and improve adaptability, their high computational complexity and the high data volume in DVS applications hinder real-time operation of the DVS systems.
Existing mitigation strategies present fundamental trade-offs: data down-sampling techniques \cite{yangRealTimeFOTDRVibration2022a,wuSmartFiberOpticDistributed2023} sacrifice signal fidelity through handcrafted preprocessing, while compressed sensing approaches \cite{shenFastStorageOptimizedCompressed2024} achieve limited compression ratios with marginal throughput improvements. 
To address these issues, knowledge distillation has emerged as a promising strategy for deploying lightweight neural networks with reduced computational demands \cite{luo2024realtimeeventrecognitionlongdistance}. However, existing implementations still fall short in meeting the requirements of large-scale, real-time systems, particularly over long distances exceeding 100 kilometers.
Moreover, conventional distillation frameworks typically operate as black-box processes and lack interpretability. Additionally, these methods often depend on high-capacity teacher models, resulting in increased training costs and complexity.

To address these limitations, our solution combines three key elements: a lightweight convolutional neural network (CNN) architecture employing depth-wise separable convolutions, FPGA-based hardware acceleration for real-time inference and a novel cross-domain knowledge distillation framework that incorporates physical priors.
We develop a three-layer depth-wise separable CNN with merely 4141 trainable parameters, which is also the most compact architecture reported so far for DVS applications. Deployed on FPGA hardware, the system achieves a record-low inference latency of 0.019 ms on a high-end (Xilinx ZCU15EG) platform and 0.031 ms on a low-end (Xilinx XC7A35T) platform for a spatial-temporal sample covering a 12.5 m fiber length over 0.256 s, enabling real-time monitoring of sensing fibers up to 168.68 km and 103.08 km, respectively. 
To enhance generalization under varying environmental conditions, we introduce a cross-domain distillation framework to systematically transfer frequency-domain insights to models operating in time domain. This approach enables joint time-frequency feature learning without increasing model complexity. By bridging domain knowledge with data-driven training, it improves recognition accuracy from from 51.93\% to 95.72\% on unseen test data, demonstrating substantial gains in both robustness and interpretability.
The proposed architecture not only delivers real-time, system-native throughput but also introduces a principled approach to embedding physical understanding into deep learning models. By establishing a direct correspondence between convolution operations and spectral decomposition, our method provides an interpretable foundation for feature learning while retaining exceptional computational efficiency.

Our contributions are summarized as follows: 
\begin{itemize} 
    \item We propose a 4141-parameter depth-wise separable CNN, representing the most compact DVS model reported to date. 

    \item We introduce a novel technique, cross-domain distillation based on a priori physical knowledge, to address the generalizability limitations of lightweight models, which also enables direct use of domain-specific knowledge.

    \item We thoroughly evaluate the proposed method and model using real-world datasets and hardware, demonstrating the system’s capability to achieve real-time processing at the theoretical maximum workload.
\end{itemize}

\section{Method}
\subsection{Architecture Design} 
To design a DVS system suitable for real-time deployment, we focus on optimizing the model architecture for efficiency and adaptability. Existing work \cite{luo2024realtimeeventrecognitionlongdistance} has demonstrated that shallow CNNs can achieve strong performance in environments statistically similar to the training data, despite having limited generalization capacity across different environments.
Building upon that foundation, this paper introduces an enhanced model architecture that further reduces complexity while improving overall efficiency. Specifically, we design an extremely compact network with only three convolutional layers, each carefully configured to balance representation capacity and computational cost. 
To further reduce the model's size and arithmetic complexity, we adopt the depthwise separable convolution technique, which is a parameter-efficient alternative to conventional convolutions \cite{Howard2017MobileNetsEC}.
It decomposes standard convolution process into two steps: channel-wise filtering (depth-wise convolution) and feature combination (point-wise convolution). 
Depth-wise convolution applies a separate filter to each input channel, meaning each filter operates on only one channel, unlike standard convolution, which combines multiple channels. 
Point-wise convolution uses a $1 \times 1$ filter to combine the features across different channels. 
\begin{equation}
\mathcal{O}(K^2 \cdot C_{in} \cdot C_{out}) \rightarrow \mathcal{O}(K^2 \cdot C_{in} + C_{in} \cdot C_{out})
\label{equ:ds_conv}
\end{equation}

\begin{figure}[ht]
\centering
\subfloat{
    \includegraphics[width=0.65\linewidth]{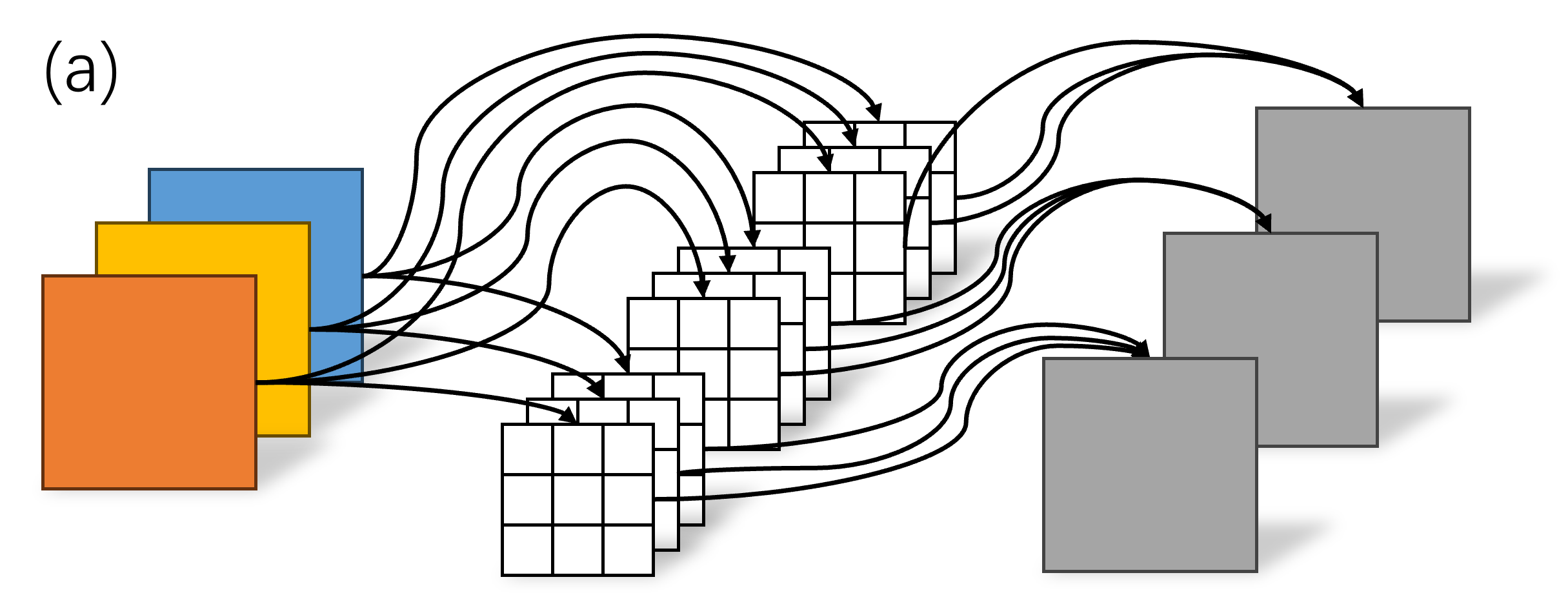}
    \label{fig:standard_conv}
}\\
\subfloat{
    \includegraphics[width=0.65\linewidth]{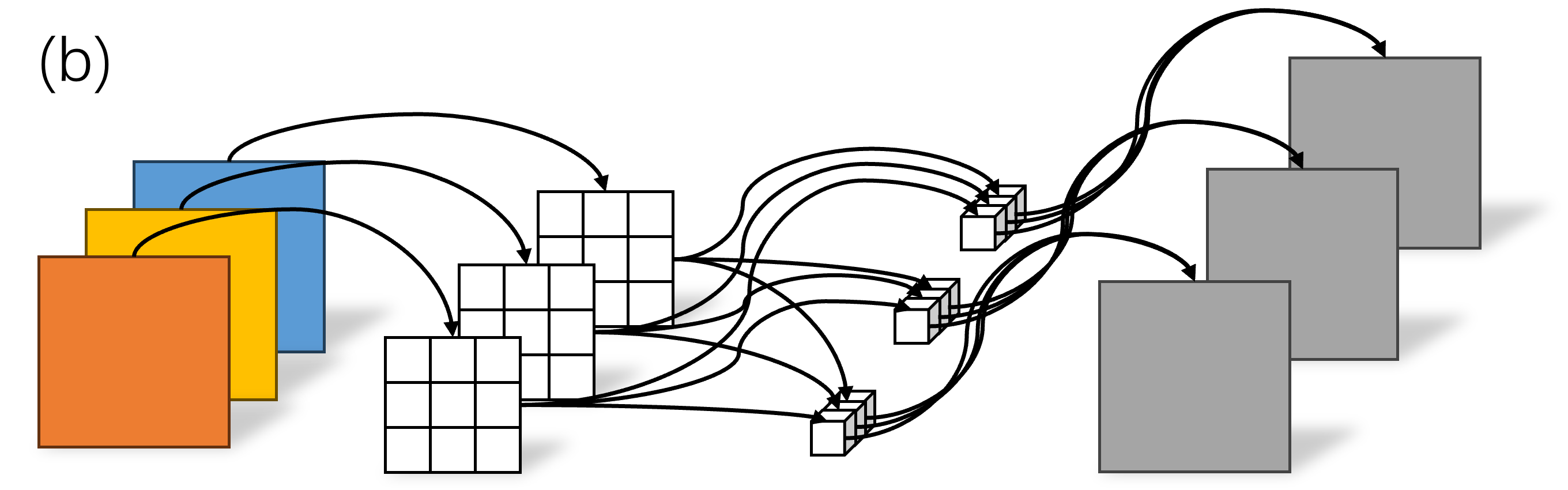}
    \label{fig:dscn}
}\\

\caption{Schematics of convolution operations: (a) Standard convolution, (b) Depth-wise separable convolution.}
\label{fig:conv_comp}
\end{figure}

This process drastically reduces the number of computations required, as shown in Eq.~\ref{equ:ds_conv}, while preserving the model’s ability to extract important features.
The process of a standard convolution is illustrated in Fig.~\ref{fig:conv_comp}(a), while the depth-wise separable convolution process is shown in Fig.~\ref{fig:conv_comp}(b). 
It is clear that the number of convolution cores involved in a depth-wise separable convolution is significantly fewer than a standard convolution.
As a result, depthwise separable convolutions significantly reduce the number of weights and arithmetic operations required, leading to substantially lower demands on computational resources and storage space

\subsection{Cross-Domain Distillation}
To address the generalizability problem of lightweight architectures in noisy vibration environments, we propose a physics-guided intra-modal distillation framework that integrates multi-domain feature learning. 
The framework strategically constrains feature learning to physically meaningful signal domains to seek for more efficient and generalizable features. 
Contrast with conventional attention-based approaches that implicitly weight features through purely data-driven parameters \cite{fengIdentificationCorrosionInducedOptical2024,yangRailwayIntrusionEvents2022,zhaoFastAccurateRecognition2022}, this framework offers inherent interpretability by directly linking learned features to selected domains investigated during training.
The framework also fundamentally diverges from cross-modality distillation techniques \cite{yeHRLiDARData2024,bangRadarDistillBoostingRadarbased}, as it operates exclusively within the vibration modality across complementary signal representations. 
This intra-modal strategy eliminates dependencies on multiple types of sensors. 
Another key advantage lies in its teacher-free design, which replaces conventional high-capacity teacher models with analytical domain references. 
This innovation reduces computational overhead drastically compared to standard distillation methods, as no teacher network requires training or inference.

\begin{figure}[ht]
    \centering
    \includegraphics[width=0.95\linewidth]{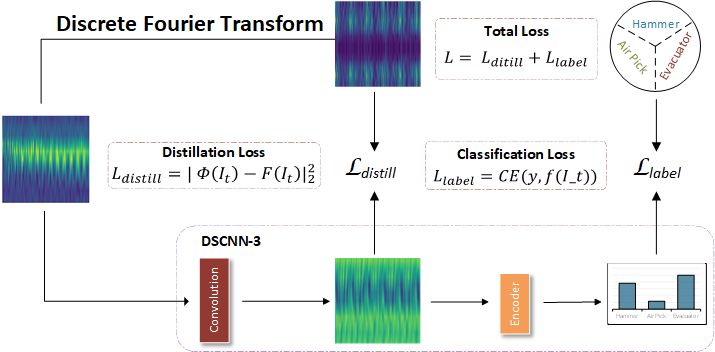}
    \caption{Schematic of Cross-Domain Distillation}
    \label{fig:kd_proc}
\end{figure}

In this paper, we adopt a simple feature-based distillation method\cite{romeroFitNetsHintsThin2015}, utilizing spectral domain for knowledge transfer.
Traditional signal processing tend to leverage Fourier analysis to extract stationary spectral features from non-stationary temporal vibration signals. 
Motivated by this physical insight, our method transfers frequency-domain knowledge to the temporal model, establishing a direct correspondence between temporal convolution operations and spectral decomposition processes, so that combining the advantages of classic signal processing with data-driven learning.
As shown in Fig.~\ref{fig:kd_proc}, our approach improves temporal learning process with spectral information, establishing a dual-domain learning paradigm.
The distillation process employs constrained optimization to enforce spectral awareness in temporal feature leaning. 
Let \( I_t \in \mathbb{R}^{T×S} \) denote the temporal input with \( T \) time steps and \( S \) spatial points. We formulate the distillation objective using the Fourier transform operator \( \mathcal{F} \):

\begin{equation}
\mathcal{L} = \underbrace{\alpha \| \Phi(I_t) - \mathcal{F}(I_t) \|_2^2}_{\text{Spectral matching}} + \underbrace{(1-\alpha)\mathcal{L}_{\text{CE}}(y, f(I_t))}_{\text{Classification}}
\label{eq: kd_proc}
\end{equation}

where $\Phi(I_t)$ represents intermediate convolutional features, $\mathcal{F}(I_t)$ denotes Discrete Fourier Transform (DFT) coefficients, $f(I_t)$ is the classification output, and $\alpha \in [0,1]$ serves as a domain weighting coefficient optimized through grid search. 
This dual-objective formulation encourages the network to learn feature representations that not only retain critical spectral characteristics but also enhance discriminative capability for accurate classification. 
By balancing these objectives, the model achieves a more robust and informative embedding space that supports both interpretability and performance.
The theoretical foundation of our approach lies in establishing a connection between depthwise separable convolutions and Fourier analysis. For an $N$-point signal $x[n]$, we demonstrate that the depthwise convolution stage can approximate the DFT matrix $F \in \mathbb{C}^{N \times N}$ through adaptive filter learning:

\begin{equation}
\underbrace{\sum_{c=1}^C (x_c * h_c)[m]}_{\text{Depthwise}} \rightarrow \underbrace{\sum_{k=0}^{N-1} x[n]e^{-j2\pi kn/N}}_{\text{DFT}}
\label{equ: csa}
\end{equation}

where $x_c$ denotes the $c$-th channel input and $h_c$ the learnable kernel for channel $c$. 
Unlike the fixed basis functions in DFT, our convolutional kernels adaptively learn frequency-selective filters through gradient descent. 
The subsequent point-wise convolution synthesizes these channel-specific spectral components into a comprehensive frequency representation, mirroring the DFT's orthogonal basis combination. 
This constrained distillation approach ensures that early convolutional layers capture both temporally localized features and their spectral signatures, providing subsequent network layers with a physically-grounded foundation for extracting noise-robust, task-specific representations.

\begin{figure}[h]
\centering
\includegraphics[width=0.85\linewidth]{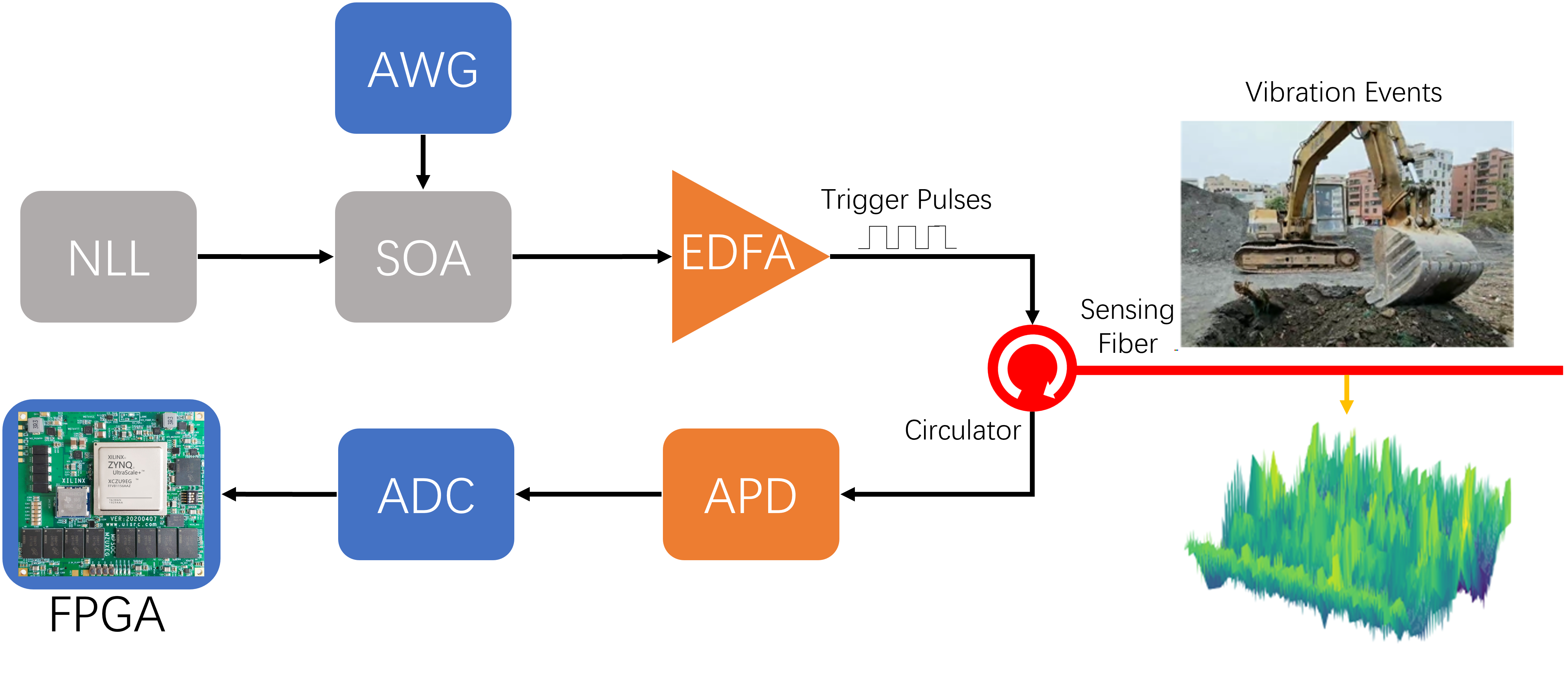}
\caption{Schematic of the proposed DVS System. A Narrow Linewidth Laser (NLL) emits continuours optical waves, shaped and amplified by an Arbitrary Waveform Generator (AWG) and a Semiconductor Optical Amplifier (SOA), then further boosted by an Erbium-Doped Fiber Amplifier (EDFA). The generated optical pulses travel through the sensing fiber, where vibration-induced Rayleigh back-scattering is detected by an Photodiode (PD), digitized by an Analog-to-Digital Converter (ADC), and analyzed in real time using the proposed model on an FPGA platform.}
\label{fig:dvs_sys}
\end{figure}

\section{Experiment}
To validate our framework, we have collected real-world DVS data using the system illustrated in Fig. \ref{fig:dvs_sys}.
The setup comprises a 30-km-long fiber, with a 50-meter long segment buried beneath 0.5 meters of composite material, which consists of soil, sand and stones in random proportions. 
This configuration introduces realistic environmental noise and variation in vibration patterns , enabling robust evaluation of model generalizability under variable conditions.

\begin{table}[h]
    \centering
    \caption{Class Label and Distribution of Samples of the Datasets}
    \begin{tabular}{ccc}
        \hline
        Class & Dataset-1 & Dataset-2\\
        \hline
        Hammer & 3332 & 268\\
        Air Pick & 3558 & 330\\
        Excavator & 3359 & 277\\
        \hline
    \end{tabular}
    \label{tab:dataset_tab}
\end{table}

\begin{figure*}[h]
\centering
\subfloat[Air Pick]{
    \centering
    \includegraphics[width=0.32\linewidth]{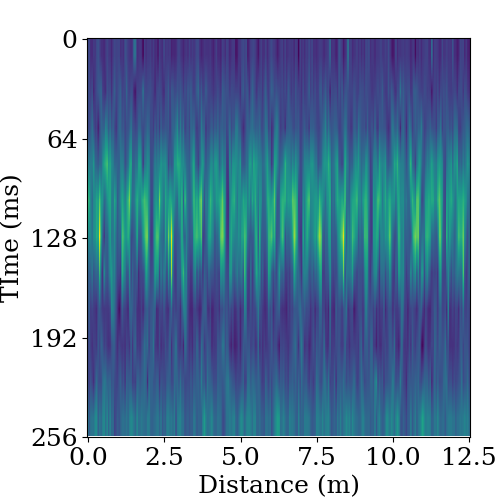}
    \label{subfig:train_air_pick}}
\subfloat[Excavator]{
    \centering
    \includegraphics[width=0.32\linewidth]{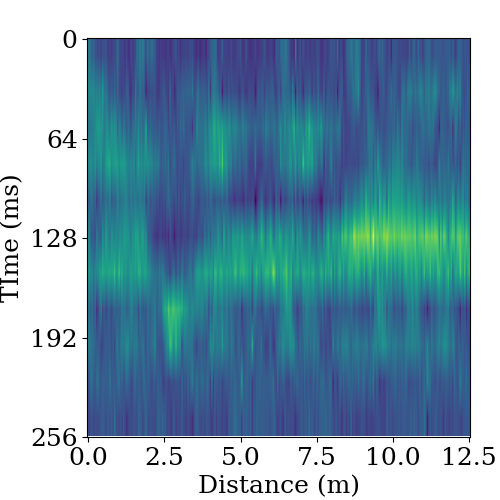}
    \label{subfig:train_excavator}}
\subfloat[Hammer]{
    \centering
    \includegraphics[width=0.32\linewidth]{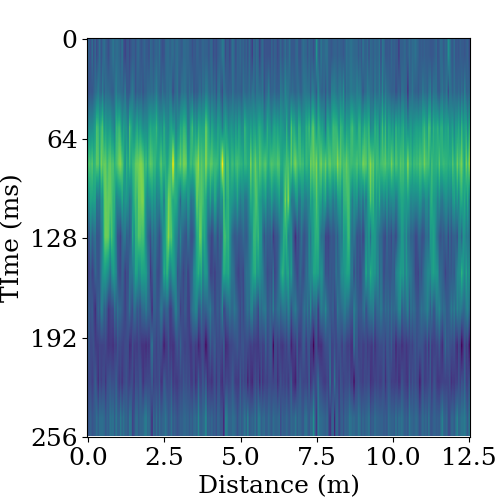}
    \label{subfig:train_hammer}}
    
\medskip

\subfloat[Air Pick]{
    \centering
    \includegraphics[width=0.32\linewidth]{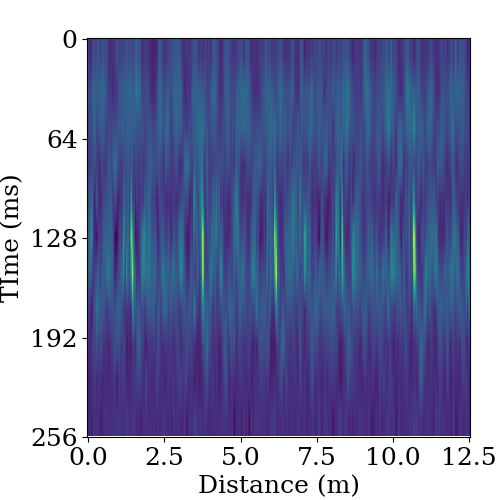}
    \label{subfig:test_air_pick}}
\subfloat[Excavator]{
    \centering
    \includegraphics[width=0.32\linewidth]{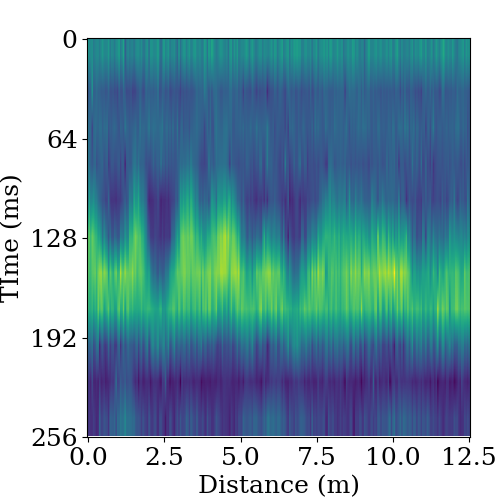}
    \label{subfig:test_excavator}}
\subfloat[Hammer]{
    \centering
    \includegraphics[width=0.32\linewidth]{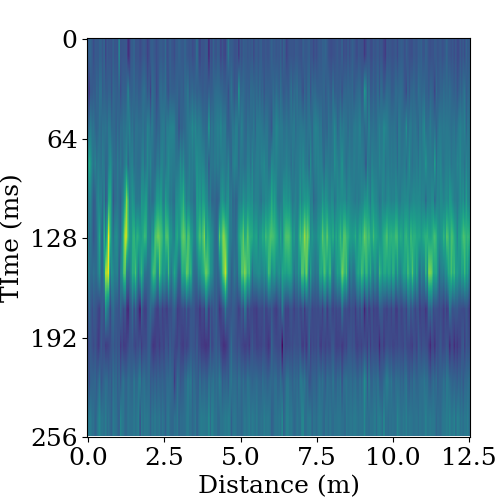}
    \label{subfig:test_hammer}}
\caption{Visualization of the Samples in Spatial-Temporal Format: (a), (b), (c) are from Dataset-1, (d), (e), (f) are from Dataset-2. Color intensity represents normalized signal amplitude.}
\label{fig:spatial_temporal}
\end{figure*}

The system captures a Rayleigh back-scattered light trace at 1-millisecond intervals, with a spatial interval of 1.5 meters. 
These traces are aggregated and segmented into samples of $256\times11$, each representing a 12.5-meter fiber segment over a 256-millisecond time window.
Data was collected at two distinct locations at different days, yielding two independent datasets.
The class distribution of these datasets is detailed in Table \ref{tab:dataset_tab}. 
Fig. \ref{fig:spatial_temporal} illustrates the inherent variability between dataset classes, demonstrating the challenge of extracting reliable signal features through lightweight architectures alone. To support cross-domain evaluation, we have generated complementary spatial-spectral representations via DFT along time axis, i.e. column direction of the samples.

\subsection{Model Optimization}

\begin{figure}[h]
\centering
\includegraphics[width=0.75\linewidth]{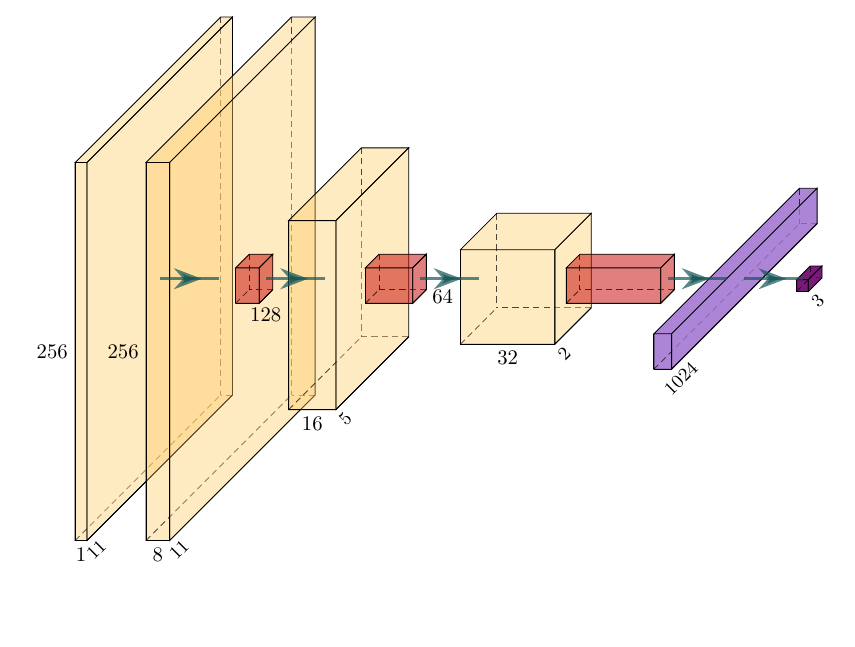}
\caption{Inference process of the DSCNN-3 model. Each box indicates the output feature map at a specific layer in the network.}
\label{fig:inf_proc}
\end{figure}

We implement a rigorous evaluation protocol using Dataset-1 for training and validation with 5-fold cross-validation and Dataset-2 for generalization testing. 
All metrics report mean values across five independent trials.
The Adam optimizer initializes with a learning rate of 0.01, reduced by half every five epochs upon validation loss plateau detection (patience=5), using a batch size of 8 for stable gradient estimation over 200 training epochs.

\begin{figure}[h]
    \centering
    \includegraphics[width=0.75\linewidth]{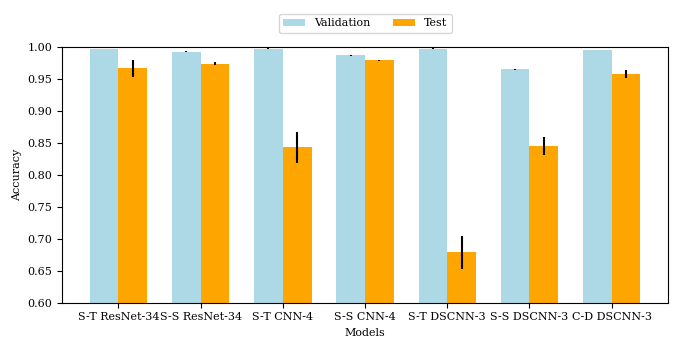}
    \caption{Performance of the DSCNN-3 and ResNet-34 models trained with Different Training Pattern. S-T: spatial-temporal model trained on raw spatial-temporal data; S-S: spatial-spectral model, trained on spatial-spectral data derived via DFT; C-D: cross-domain model, trained using cross-domain distillation.}
    \label{fig:model_res}
\end{figure}

Based on previous works, we develop a more compact 3-layer CNN architecture (DSCNN-3) that incorporates depth-wise separable convolutions to reduce complexity while maintaining performance. 
We further implement a popular large size network ResNet-34 for comparison\cite{Jin2023PatternRO, Yao2023VibrationER}.
DSCNN-3 consists of three depth-wise separable convolutional layers, each followed by pooling layers: the first two use max pooling to down-sample the feature maps, while the third uses average pooling to preserve more contextual information. 
These layers gradually increase the number of output channels from 8 to 16 and finally 32. 
After the convolutional blocks, the output is flattened and passed to a fully connected layer that maps features to the final class predictions.
Fig. \ref{fig:inf_proc} illustrates the end-to-end inference pipeline, including the spatial–temporal input, the DFT-guided first convolution, which introduces spectral sensitivity, subsequent convolution layers and pooling layers that extract hybrid spatial–spectral features, and the final classification layer.

To quantify the impact of the cross-domain distillation framework, we compare the models trained with three different training paradigms: the S-T model, trained on raw spatial-temporal data; the S-S model, trained on spatial-spectral data derived via DFT; and the C-D model, trained using cross-domain distillation.
As shown in Fig.~\ref{fig:model_res}, DSCNN-3 matches the validation accuracy of both CNN-4 and ResNet-34, despite having fewer parameters. It confirms the efficacy of depth-wise separable convolutions for preserving model performance while reducing model complexity. 
But the accuracies of S-S and S-T DSCNN-3 shows a steep drop on the test set, illustrating the limited generalizability of the model.
Fig.~\ref{fig:model_res} also visualizes the comparative performance of the three training paradigms.  The distilled DSCNN-3 not only matches the deep ResNet-34 in overall accuracy, but also narrows the gap between validation and test performance, indicating improvements in generalizability and robustness.  By contrast, the S-T model, while achieving high validation accuracy, suffers a steep drop in test accuracy, and the S-S model yields moderate generalizability. The cross-domain distillation framework guide the model to search for features within two domains, thus getting both efficient and robust features, which produce a better performance and generalizability.

\begin{figure*}[h!]
\centering

\subfloat[S-T Model]{
   \includegraphics[width=0.32\linewidth]{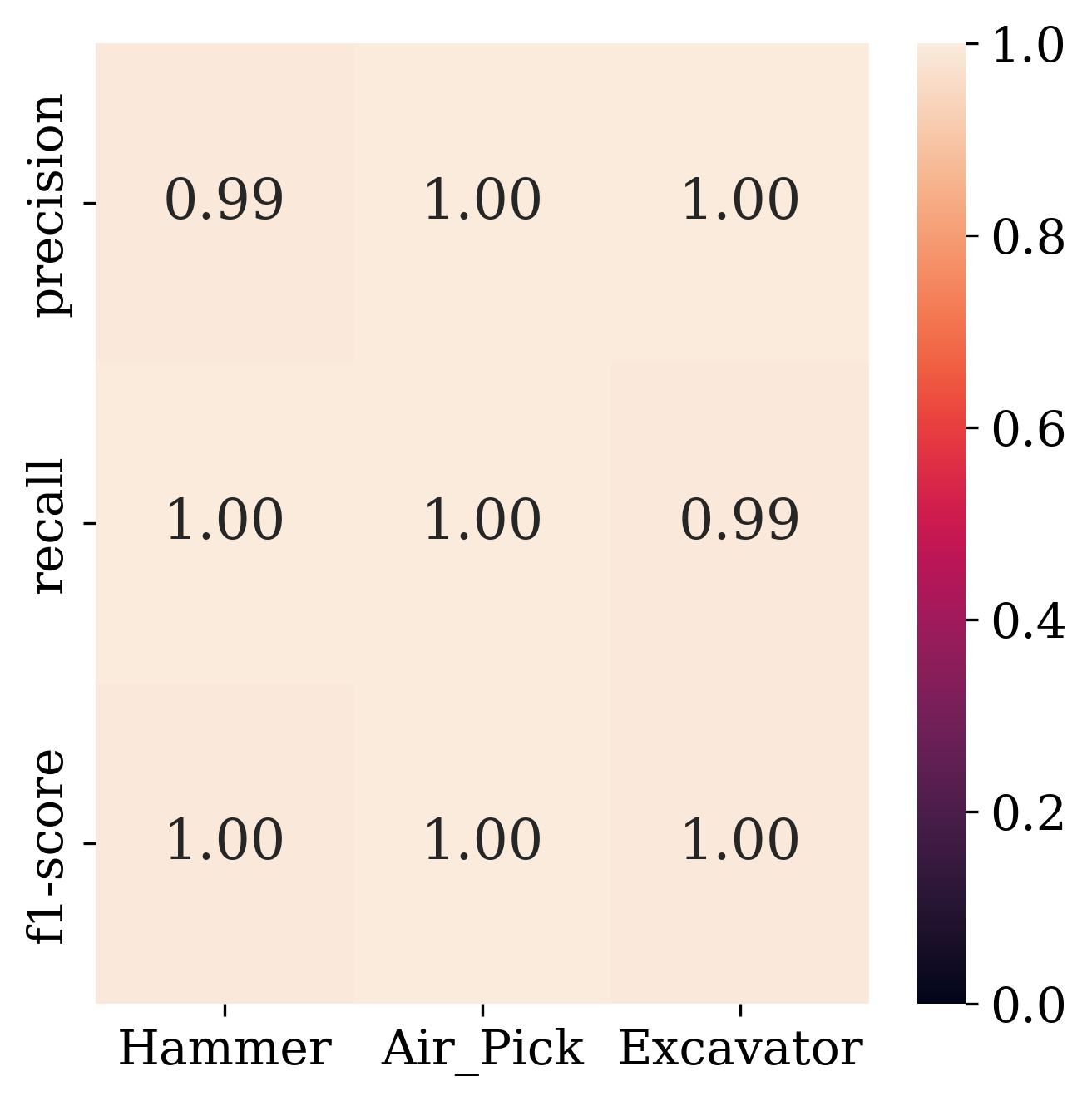}
}
\subfloat[S-S Model]{
   \includegraphics[width=0.32\linewidth]{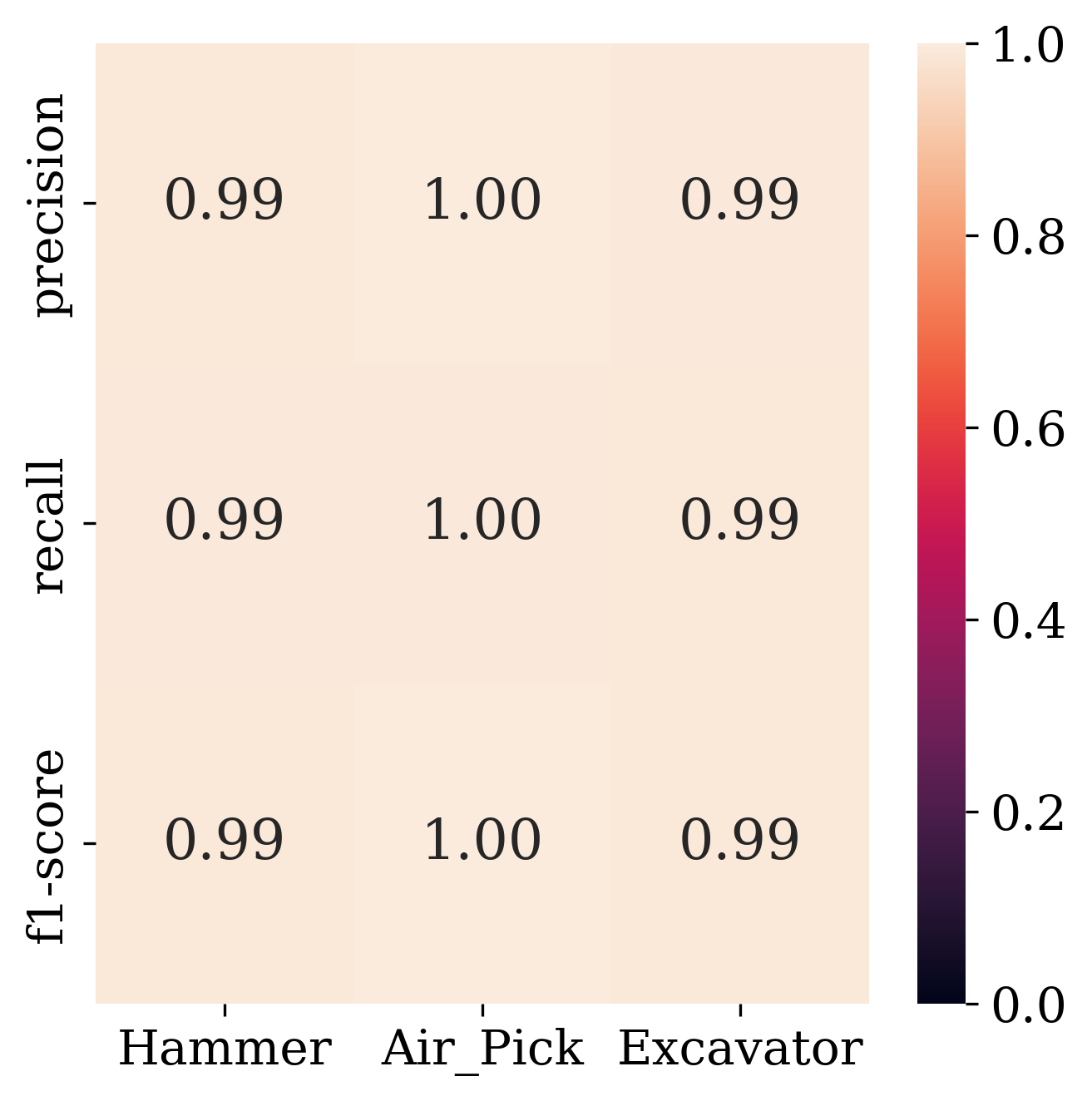}
}
\subfloat[C-D Model]{
   \includegraphics[width=0.32\linewidth]{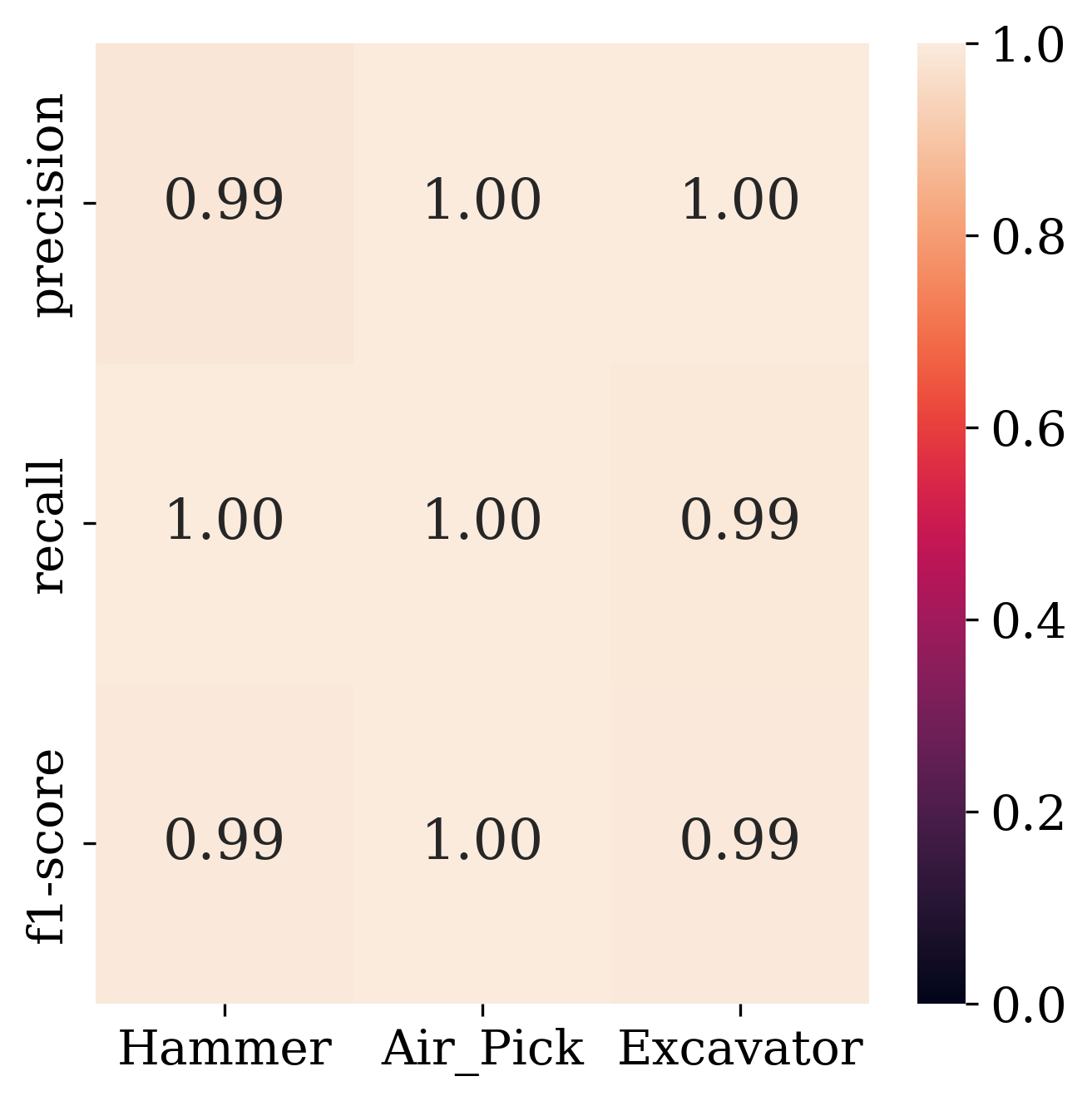}
}

\medskip

\subfloat[S-T Model]{
   \includegraphics[width=0.32\linewidth]{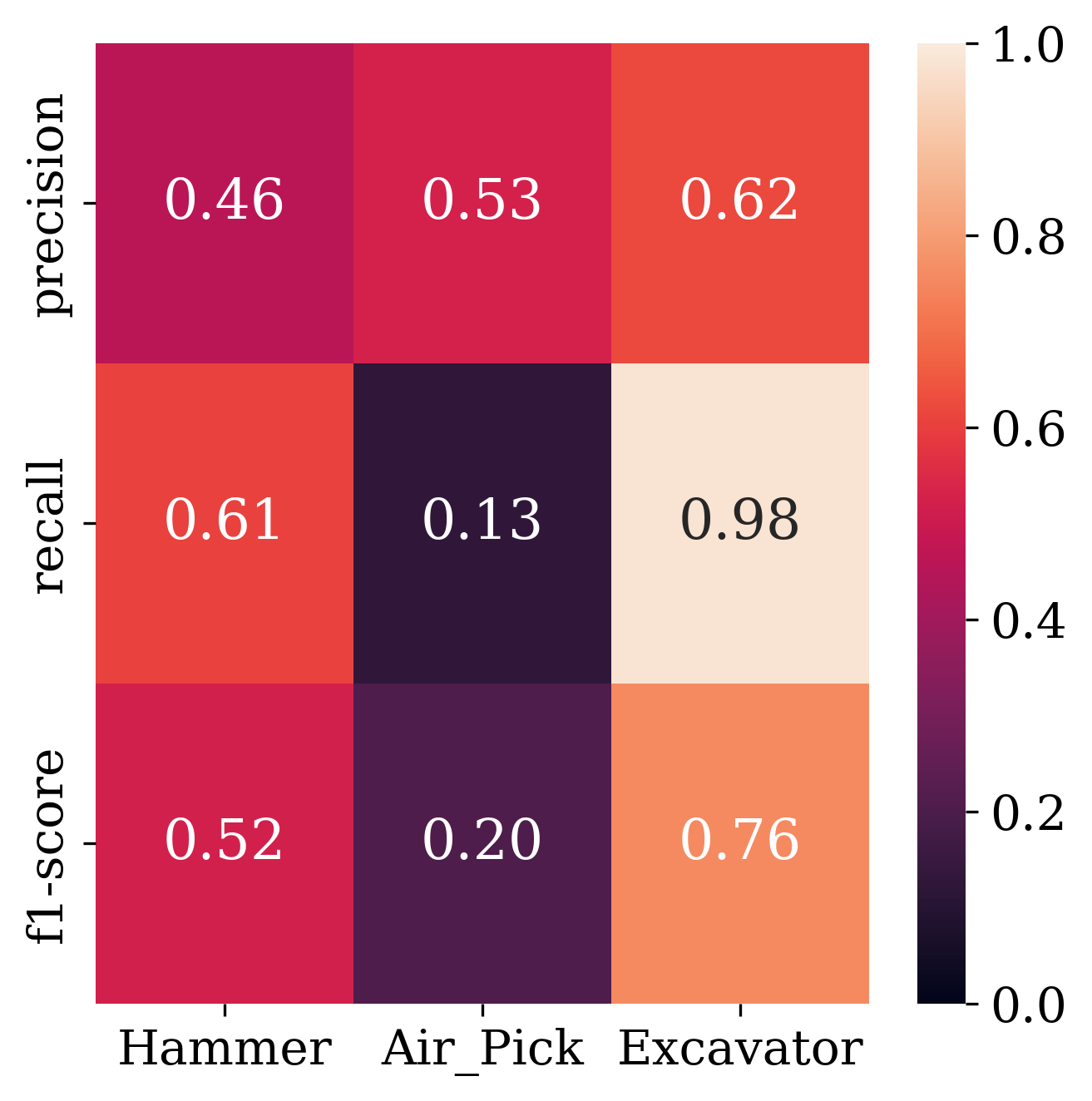}
}
\subfloat[S-S Model]{
   \includegraphics[width=0.32\linewidth]{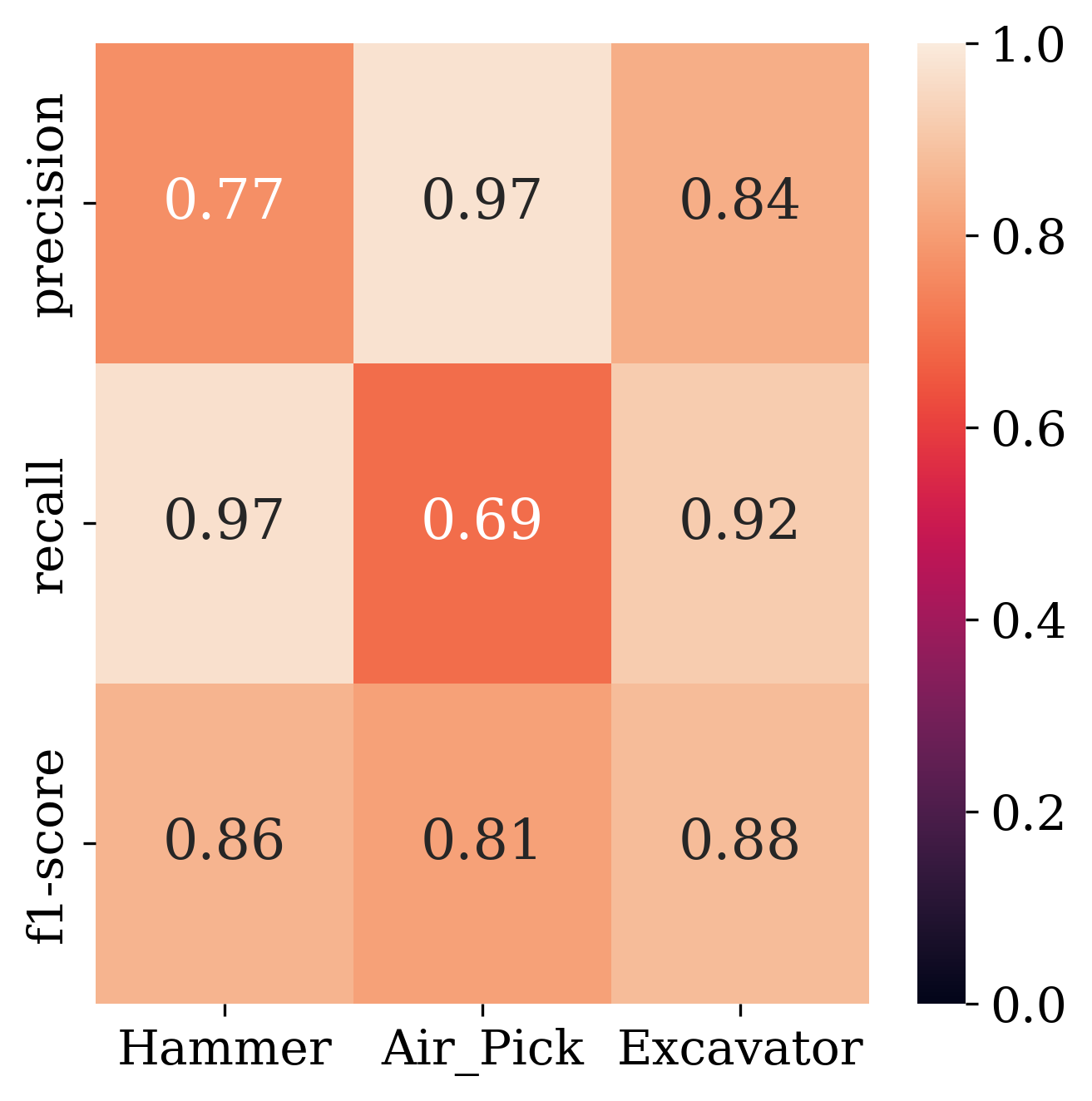}
}
\subfloat[C-D Model]{
   \includegraphics[width=0.32\linewidth]{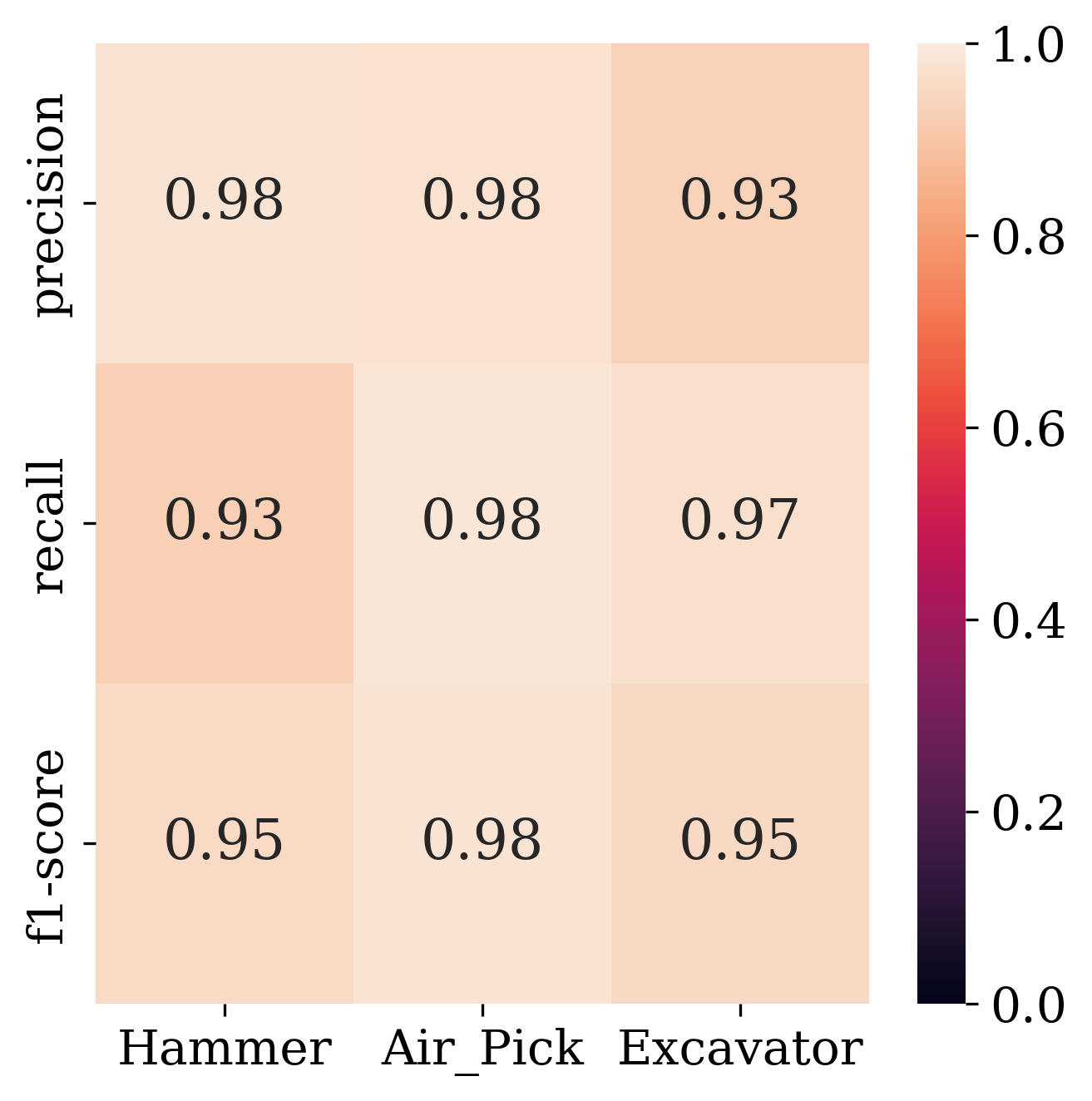}
}
\caption{Visualization of the Results of the Models: (a), (b), (c) are the validation results, (d), (e), (f) are the test results.}
\label{fig:cm_results}
\end{figure*}

Fig.~\ref{fig:cm_results} offers deeper insight into these trends. In the S-T model result on the test set, low precision and recall reveals its over-reliance on local temporal patterns. The S-S model mitigates some of these errors but still exhibits confusions, as shown in the low precision and high recall in "Hammer" class, and high recall but low precision in "Air Pick" class.
In contrast, the C-D model shows a much more balanced performance matrix.  This demonstrates that cross-domain distillation not only lifts overall accuracy but, critically, distributes that accuracy evenly across all categories, validating its role in guiding the lightweight network to learn both efficient temporal features and generalizable spectral signatures.

\begin{figure}[ht]
    \centering
\subfloat[Validation Accuracy vs. Depth]{
   \includegraphics[width=0.45\linewidth]{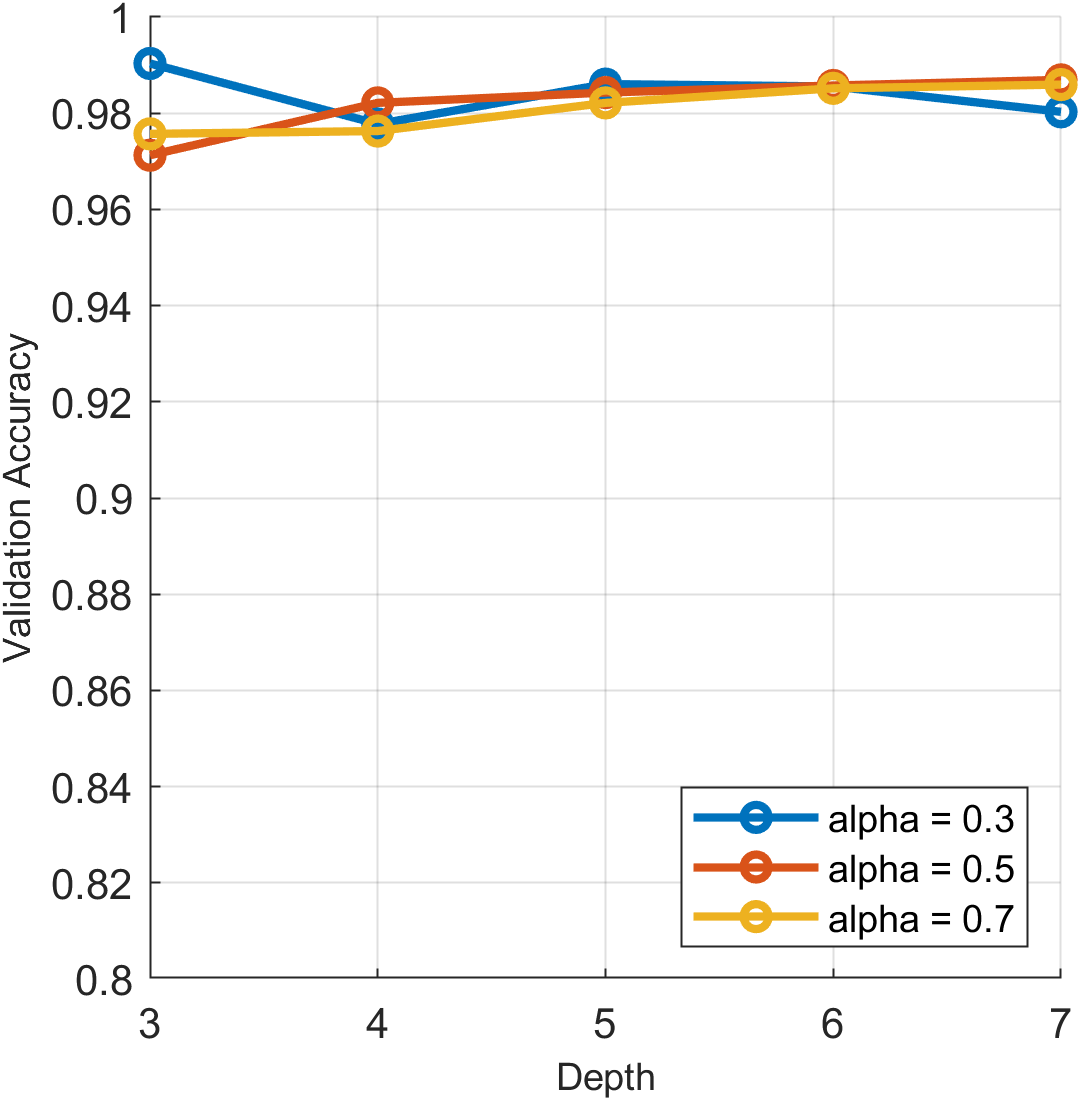}
}
\subfloat[Test Accuracy vs. Depth]{
   \includegraphics[width=0.45\linewidth]{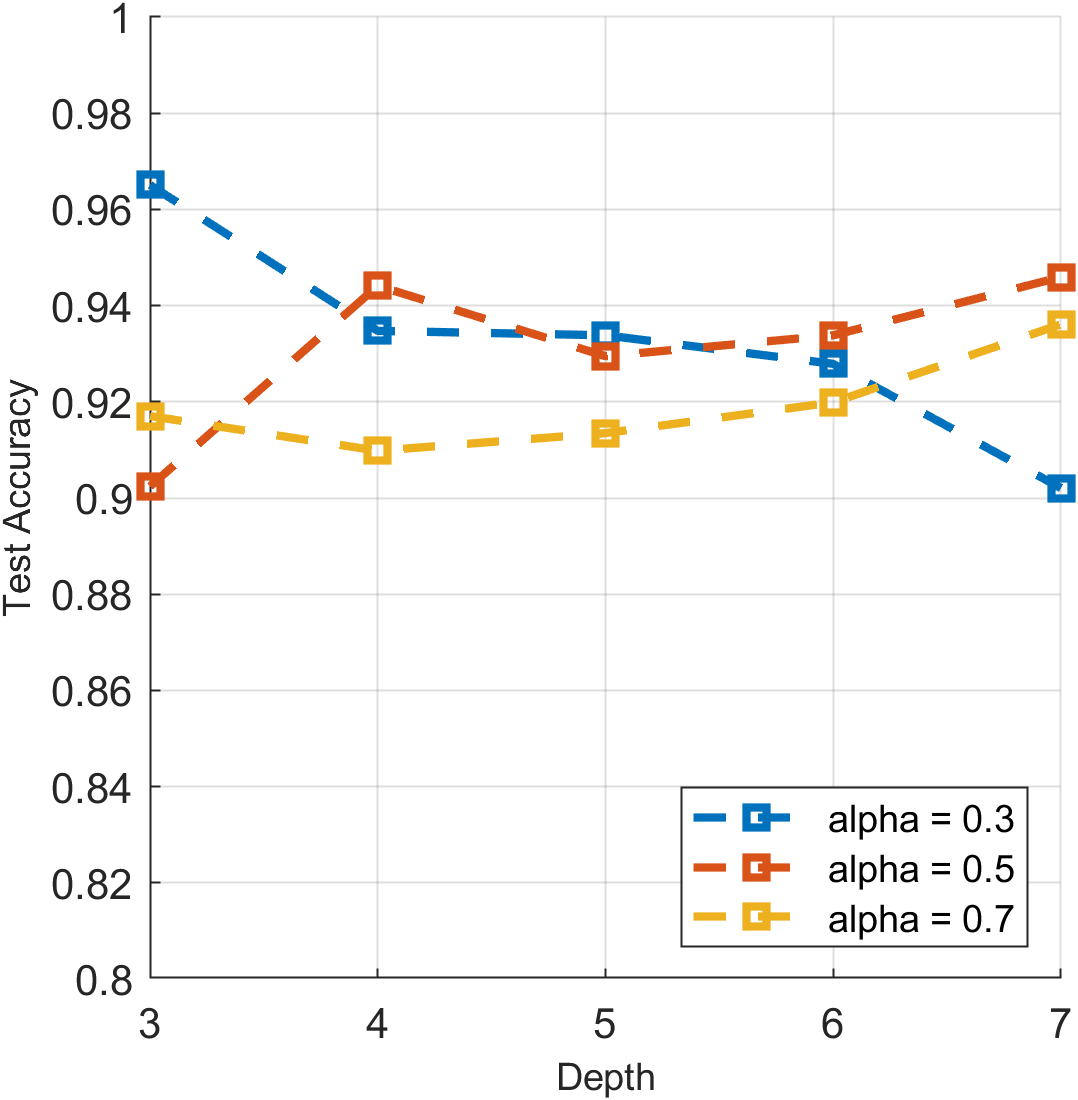}
}
        \caption{
        Model performance across network depths and $\alpha$ values with cross-domain distillation: (a) is the validation accuracies, (b) is the test accuracies.}
    \label{fig:avd}
\end{figure}

To investigate depth-impact on distillation efficacy, we further conduct controlled experiments across 1-5 convolutional layers preceding the pooling layers, with three $\alpha$ values (0.3, 0.5, 0.7).
This is also aimed to investigate the question that, while deep layers does help provide better capacity, but does this improved network capacity help the knowledge distillation process, or help to better approximate the DFT process to introduce spectral information to the low-level features, while extracting temporal information.
Fig.~\ref{fig:avd} plots both validation and test accuracies for different depth and $\alpha$ values.  The one-layer variant achieves a test accuracy of 95.72\%, slightly higher than that of deeper models and very close to the best ResNet-34 baseline.  
Beyond a single convolution layer, additional layers yield diminishing returns: while minor gains appear with deeper architecture, these do not shows improvement compared to the single-layer model, thus does not provide any reason for using deeper architecture. 
This plateau suggests that the core spectral-temporal alignment enforced by cross-domain distillation is already captured by the first convolution layer, in line with our theoretical insight that a depthwise separable convolution can effectively approximate the linear DFT mapping (Eq. \ref{equ: csa}).
These results have two important implications.  First, they confirm that an appropriately structured, minimal network suffices to absorb the knowledge transfer from another domain, alleviating the need for deeper and more computationally expensive architectures.  
Second, they highlight that the primary benefit of adding layers is not to improve distillation effect, but to refine hierarchical feature abstractions, which, in our lightweight setting, proves unnecessary for this specific cross-domain task.  Together, these findings validate our design choice of a single-layer distilled CNN, striking the optimal balance between parameter efficiency, spectral alignment, and temporal feature learning.

\subsection{Hardware Acceleration}
To optimize the computation speed of the proposed model, we implement it on an FPGA platform, leveraging its reconfigurability and flexibility. 
To further accelerate processing, we employ a specialized "shift-add" quantization scheme\cite{luo2024realtimeeventrecognitionlongdistance}. 
In this scheme, traditional resource-intensive multiplication operations are replaced by shift operations, enabling computations to be handled by the FPGA's logic units rather than its limited digital signal processing (DSP) resources \cite{Meng2021FixyFPGAEF}. 
This transformation significantly enhances parallelism, allowing more operations to be executed simultaneously, thereby reducing latency.

For practical assessment, the high-end Xilinx ZCU15EG platform is chosen for performance evaluation, as it offers abundant resources for full parallelization and clock rate optimization. 
This implementation achieves an impressive 8,469-cycle latency at a 2.24 ns clock period, enabling real-time processing of 13,494 samples per second, with consumption of only 2.55\% of available LUTs (8,690 out of 341,280) and 1.32\% of FFs (9,020 out of 682,560). Remarkably, it consumes no DSP blocks, BRAM or URAM, making it highly suitable for deployment on cost-sensitive and resource-constrained edge devices.
This processing speed supports the monitoring of 168.7 km of optical fiber at a 1.25 m spatial interval. 

\begin{table*}[!ht]
    \centering
    \caption{Computational Efficiency and Model Performance Comparison between Distilled Models and Deep Models}
    \begin{tabular}{ccccccc}
    \hline
        Model & Inference time & Inference time & Val/\% & Test/\% & FLOPs & Params\\
         & on GPU/ms & on FPGA/ms &  &  &  & \\
    \hline
        ResNet-34 & 0.375 & - & 99.50 & 97.67 & 226235392 & 8164803\\
        CNN-4\cite{luo2024realtimeeventrecognitionlongdistance} & 0.178 & 0.083 & 99.61 & 95.39 & 4331520 & 30771\\ 
        DSCNN-3 & 0.162 & 0.019 & 99.49 & 95.73 & 601600 &  4141\\
    \hline
    \end{tabular}
    \label{tab:inf_time}
\end{table*}

As shown in Table~\ref{tab:inf_time}, the distilled DSCNN-3 model significantly outperforms both the traditional deep model and the previously published lightweight model in terms of computational efficiency while maintaining competitive accuracy. DSCNN-3 accomplishes this with a drastic reduction in computational complexity compared ResNet-34. This highlights the effectiveness of our cross-domain distillation and lightweight architectural design.
In terms of inference latency, DSCNN-3 demonstrates substantial speed advantages. On GPU, it outperforms the lightweight CNN-4 model and significantly faster than ResNet-34.
More importantly, the FPGA implementation of DSCNN-3 delivers an ultra-low latency of only 0.019 ms, which is 4.37× faster than the 0.083 ms achieved by CNN-4 using the same implementation scheme and FPGA platform.
This improvement is primarily attributed to the efficient use of depthwise separable convolutions and the shift-add quantization strategy, both of which enable high degrees of parallelism and optimized resource utilization on the FPGA.
Notably, ResNet-34 is not implemented on FPGA due to its high resource requirements, which make full parallelization unfeasible on practical hardware. 
The contrast between the GPU and FPGA inference times further reveals efficient use of hardware resources of our model and implementation scheme, especially considering the maximum clock speed of used GPU is at 2520 MHz, which is about 5 times faster then the FPGA implementation.
While deep models often benefit from GPU acceleration, they cannot take full advantage of FPGA's parallel processing capabilities due to their high parameter and computation complexity. 
In contrast, the DSCNN-3's streamlined design and quantized arithmetic are tailored for FPGA implementation, resulting in real-time throughput and minimal latency without sacrificing accuracy.

To assess deployability in resource-constrained environments, we also implement DSCNN-3 on a low-end Xilinx XC7A35T platform.
This implementation achieves real-time compliance with 103.1 km monitoring capability with clock period relaxed to 4.9 ns due the limited hardware resources. 
The proposed design demonstrates exceptional hardware efficiency, utilizing only 8,579 look-up tables (LUTs) and 10,068 flip-flops (FFs) on the low-end XC7A35T FPGA, equivalent to just 20.6\% and 48.4\% of available resources, respectively, again with zero DSP, BRAM, or URAM consumption. 

Both implementations exceed the 100 km sensing range, which aligns with the theoretical maximum sensing range of the evaluation system used in this paper. 
Overall, the above results clearly demonstrate that the proposed DSCNN-3 model, along with proposed methods, offers a well-balanced trade-off between accuracy, computational cost, and hardware efficiency. It serves as a practical and scalable solution for real-time DVS applications, especially in resource-constrained environments where conventional deep models are infeasible to deploy. 

\section{Conclusion}
This paper addresses the critical challenge of real-time processing in DVS systems, which has traditionally been limited by both hardware constraints and the heavy computational load of conventional deep learning models.
The proposed framework combines an ultra-lightweight neural network (DSCNN-3) with FPGA-based acceleration to deliver high-speed inference even on resource-limited platforms.
The proposed depth-wise separable CNN (DSCNN-3) achieves state-of-the-art performance with only 4,141 parameters, reducing model complexity by 99.95\% compared to ResNet-34 while maintaining similar performance and generalizability.
Combined with a shift-add quantization scheme optimized for FPGA implementation, our solution achieves implementation on commercial grade FPGA chips while providing process capability of the maximum theoretical sensing range of the system, demonstrating high applicability in real-world applications.
This paper proposes the cross-domain distillation technique to improve generalizability of lightweight models, and to introduce physical knowledge into the training process, to provide more interpretability.
By integrating signal processing techniques with data-driven learning, this method embeds prior physical knowledge, such as spectral-temporal relationships derived from Fourier analysis, directly into modern neural networks.
Our experiments demonstrate that cross-domain distillation improves test accuracy by 43.8\% compared to the baseline model trained with the conventional method, addressing the critical issue of generalizability of lightweight models in various environments.
By unifying lightweight architectures, physics-informed distillation, and hardware acceleration, this work provides a scalable and efficient solution for real-time, long-distance vibration monitoring. 
Beyond DVS, the proposed approach provides a pathway for integrating domain-specific physical knowledge into neural networks, offering a generalizable strategy for building robust, efficient and explainable deep-learning-based systems in DOFS area.

\begin{backmatter}
\bmsection{Funding}
National Natural Science Foundation of China (62225110, 61931010); the Major Program (JD) of Hubei Province (2023BAA013); Hubei Provincial Natural Science Foundation of China (2025AFB008).

\bmsection{Disclosures} The authors declare no conflicts of interest.

\bmsection{Data availability} Both data and codes underlying the results presented in this paper are available in GitHub repository https://github.com/HUST-IOF/Cross-Domain\_Distill.

\end{backmatter}

\bibliography{KD-FFT}

\end{document}